\documentclass[useAMS,usenatbib]{mn2e}
\usepackage{amssymb}
\usepackage[dvips]{graphicx}

\title[The chemistry of transient microstructure in the diffuse ISM]{The chemistry of transient microstructure in the diffuse interstellar medium}
\author[T. A. Bell, S. Viti, D. A. Williams, I. A. Crawford and R. J. Price]
{T. A. Bell,$^{1}$\thanks{E-mail: tab@star.ucl.ac.uk} S. Viti,$^{1}$ D. A. Williams,$^{1}$ I. A. Crawford$^{2}$ and R. J. Price$^{1}$\\
$^{1}$Department of Physics \& Astronomy, University College London, Gower Street, London WC1E 6BT\\
$^{2}$School of Earth Sciences, Birkbeck College, Malet Street, London WC1E 7HX}
\begin{document}

\date{\today}
%\date{Accepted 2004 December 1. Received 2004 October 5; in original form 2004 April 16}

%\pagerange{\pageref{firstpage}--\pageref{lastpage}} \pubyear{2004}

\maketitle

\label{firstpage}

\begin{abstract}
Transient microstructure in the diffuse interstellar medium has been
observed towards galactic and extragalactic sources for decades,
usually in lines of atoms and ions, and, more recently, in molecular
lines. Evidently, there is a molecular component to the transient
microstructure. In this paper we explore the chemistry that may arise
in such microstructure. We use a PDR code to model the conditions of
relatively high density, low temperature, very low visual extinction
and very short elapsed time that are appropriate for these objects. We
find that there is a well defined region of parameter space where
detectable abundances of molecular species might be found. The best
matching models are those where the interstellar microstructure is
young ($<100$ yr), small ($\sim100$ AU), and dense ($>10^4$
cm$^{-3}$).
\end{abstract}


\section{Introduction}
Observational evidence for microstructure on a solar-system scale in the 
diffuse interstellar medium has been accumulating for the last three 
decades. Measurements include VLBI of H atom 21 cm absorption toward 
extragalactic sources \citep{b12,b10,b9}, absorption of atomic lines 
toward transversely moving pulsars \citep{b13}, absorption differences 
toward stellar binaries \citep{b21,b20} and toward multiple systems and 
clusters \citep{b19,b18,b17}, and secular variations over intervals of a 
few years along the line of sight to single stars \citep{b16,b7,b3,b23,
b26,b1,b8,b6}. \citet{b30} find secular changes in formaldehyde
absorption towards extragalactic sources, the first report of
molecular tracers of microstructure. \citet{b31} infer structure on
scales of order 10 AU, with number densities possibly larger than
10$^6$ cm$^{-3}$, from further formaldehyde observations toward
extragalactic sources. They also find spatial variations in
formaldehyde and hydroxyl absorption line profiles towards the
extended radio galaxy 3C 111. \citet{b29} and \citet{b28} have used
similar observations to explore the range of chemistry in diffuse and
translucent clouds. It is also well known \citep{b27} that structure
at larger scales ($\sim0.01$ pc) exists in nearby clouds.

\citet{b5} has recently detected absorption in lines of CH at the
specific velocity (different to that of the surrounding
medium) of a previously reported variable component in the line of
sight to $\kappa$ Vel. Using a simple chemical model, Crawford
obtained a value for the number density of the CH-containing
microstructure toward $\kappa$ Vel that was consistent with the
density inferred from measurements of the Ca\textsc{I}/Ca\textsc{II}
ratio at the same velocity component:
$n_{\mathrm{H}}\sim10^{3}$ cm$^{-3}$. The measured CH column density
would require the path-length to be more than one order of magnitude
larger than the assumed transverse size, supporting the view that the
structure may be extended in the line of sight, either as a filament
or a sheet.

The most detailed study of molecular microstructure to date has been
made by \citet{b32}. These authors made spectroscopic observations of
the runaway reddened star HD 34078 (whose velocity transverse to the
line of sight is $\sim100$ km s$^{-1}$) over a 3-year period,
supplemented by data from earlier epochs, to probe the foreground
cloud distributions of CH, CH$^+$, CN and DIB carriers on scales from
1 -- 150 AU. Their results show that the CH column density increased
by 20\% in 10 years while the CH$^+$ column density and two DIB
strengths were unchanged. The CN column density shows a modest rise in
this period.

Although some authors have argued that some of the radio evidence for
microstructure is flawed by poor data handling (e.g.~Stanimirovic et
al.~2003), the evidence in favour of the existence of microstructure
now seems overwhelming. The general conclusion is that microstructure
has a scale of 10 -- 100 AU, and has typical densities much larger
than the ambient interstellar density. The transverse visual
extinction is therefore very small, and the gas in the structure can
only be poorly shielded from the interstellar radiation field.

The origin of the microstructure has been the subject of intense study
\citep{b35,b34}. Some authors argue that it cannot originate through
non-magnetic hydrodynamics and it has been suggested that it may be
excited in regions of high magnetic pressure by slow-mode magnetosonic
waves \citep{b15}. However it is formed, it is overpressured and must
be transient. \citet{b32} inferred from their observations that the
variations were not the result of dense clumps passing through the
line of sight. We are inclined to agree with their conclusion on
dynamical grounds, and emphasise that the transience of the
microstructure implied by the high overpressure must affect the
chemistry. It is unclear whether these structures are extended,
i.e. filamentary or sheet-like, or whether they are compact
objects. Extended objects could have longer path-lengths along
the line of sight, if the orientation is suitable, and therefore the
constraints on density could be relaxed as compared to compact
objects. The geometry may be related to the origin of these objects.

The purpose of this paper is to explore the chemistry that may arise
in the conditions apparently appropriate for the transient
microstructure: rapid transition from relatively low to relatively
high density, low temperature, very low transverse visual
extinction, and subject to the normal interstellar radiation and
particle fields. In Section 2 we describe the model in more detail,
and present in Section 3 our predictions of detectable chemistry as
functions of time and depth into the slab. In Section 4 we describe
the region of parameter space in which detectable molecular column
densities might be found, and discuss the relevance of possible
detections to the geometry of the structure.

\section{The Model}
\begin{figure}
 \caption{Model geometry of microstructure at the velocity of the
 variable component in the line of sight to $\kappa$ Vel (different
 from that of the surrounding medium) with a depth transverse to the
 line of sight of 10 AU and an elongated length along the line of
 sight of 100 AU. The interstellar radiation field striking the
 microstructure transversely to the line of sight drives the chemistry
 in the microstructure, its intensity falling off with
 $A_{V,trans}$. The contribution to column densities from species
 within the microstructure is calculated along the line of sight
 through the structure, over a visual extinction $A_{V,long}$.}
 \label{geometry}
 \begin{center}
  \includegraphics[width=0.45\textwidth]{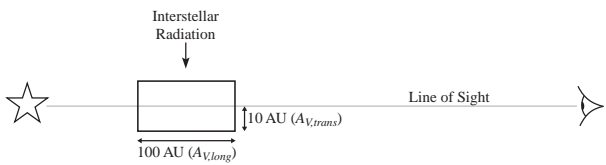}
 \end{center}
\end{figure}

We model the region crudely as a semi-infinite slab illuminated from
one side; the density rises as a step function at the edge of the
slab. We compute the chemistry as a function of time and
transverse depth position within the slab, restricting our
attention to the edge region (see Figure~\ref{geometry}). 
What is unusual about this calculation compared to other studies of
diffuse clouds is that the region of interest into the slab
transversely to the line of sight is restricted to $\sim10$ AU and to
a very short evolutionary age, typically less than 10$^3$ yr. These
tiny spatial and time domains are usually ignored in studies of
diffuse clouds (but note the models proposed by \citet{b2} to account
for H$_{3}^{+}$ and other species in the line of sight towards Cygnus
OB2). We have therefore had to develop a chemical code in which
particular attention is given to early time and edge effects. This
code is a development of the photodissociation region (PDR) code
written by Viti and Thi \citep{b22}.

We investigate the chemistry in transient microstructure using a PDR 
code that self-consistently determines the chemistry and thermal 
balance on a one-dimensional adaptive spatial grid (Papadopoulos et 
al. 2002; to be described in detail in a subsequent paper). While 
the physical structure of the slab is assumed fixed, the chemistry 
within it will be strongly time- and space-dependent. We have chosen
gas phase elemental abundances, relative to $n_{\mathrm{H}}$, as follows: 
$\mathrm{He/H}=0.075$, $\mathrm{C/H}=1.79$$\times$10$^{-4}$,
$\mathrm{N/H}=8.52$$\times$10$^{-5}$, $\mathrm{O/H}=4.45$$\times$10$^{-4}$,
$\mathrm{Na/H}=8.84$$\times$10$^{-7}$, $\mathrm{Mg/H}=5.12$$\times$10$^{-6}$,
$\mathrm{Si/H}=8.21$$\times$10$^{-7}$, $\mathrm{S/H}=1.43$$\times$10$^{-6}$,
$\mathrm{Cl/H}=1.1 $$\times$10$^{-7}$, $\mathrm{Ca/H}=5.72$$\times$10$^{-10}$,
$\mathrm{Fe/H}=6.19$$\times$10$^{-6}$. The fraction of hydrogen
initially contained in H$_2$ is unknown; the chemistry will depend
strongly on this factor. We note that the structures are dense and, if
set up as proposed by Hartquist et al. (2003), would be found in
post-shock regions. Hence, it is plausible to assume a significant
fraction of H$_2$ in the slab. We have arbitrarily set
$n(\mathrm{H_2})/n_{\mathrm{H}}=0.4$, and we note that the chemical
abundances predicted by the model should be proportional to this
parameter. Since we are considering the \textit{additional
contribution} to molecular species produced by chemistry in the
microstructure, all other molecular abundances are set to zero
initially. A total of 128 species are included in the model, connected
through a network of over 1500 reactions, including ionisation and
recombination of atoms. Freeze-out and mantle evaporation processes
have been neglected due to the short timescales over which these
structures evolve and their low visual extinction.

The filamentary or sheet-like nature of such objects is
accounted for by assuming that the line of sight is within the slab
and at a transverse depth of 10 AU from its surface
(see Figure~\ref{geometry}). Column densities are then simply
proportional to the length of the line of sight within the slab.

Since the physical conditions within these regions are uncertain, a
large range of parameter space is examined to provide a thorough
investigation of microstructure chemistry. The effect of varying the
slab density and age, as well as the environmental parameters of
incident radiation field strength and cosmic ray ionisation rate
($\zeta$) are all considered, producing a four-dimensional grid of
test parameters.

Observational estimates of microstructure density generally vary
between $10^3 \le n_{\mathrm{H}} \le 10^5$ cm$^{-3}$, with some
indications of even higher densities. We have explored the model
chemistry for this density range, and have made some additional
calculations for the higher densities suggested by the formaldehyde
observations of \citet{b31}. Slab ages of 1 -- 1000 yr are
considered. Radiation strengths and cosmic ray ionisation rates of
$1/3$, 1 and 3 times the standard interstellar values (1 Habing and
1.3$\times$10$^{-17}$ s$^{-1}$, respectively) are also considered.

\section{Model Results}
\begin{table*}
 \caption{Column densities (cm$^{-2}$) for molecular species along the
 line of sight within a microstructure of transverse depth 10
 AU and elongation factor 10 along the line of sight. The region is
 subject to standard interstellar radiation and particle fields of 1
 Habing (1.6$\times$10$^{-3}$ erg cm$^{-2}$ s$^{-1}$) and
 $\zeta=1.3$$\times$10$^{-17}$ s$^{-1}$. Values in bold indicate
 potentially detectable abundances (with column densities $>10^{11}$
 cm$^{-2}$).}
 \label{coldens1}
 \begin{center}
  \begin{tabular}{ccccccc}
\hline
Species & Time    & \multicolumn{5}{c}{Density (cm$^{-3}$)}
\\
        & (years) & $10^{3}$ & $5\times10^{3}$ & $10^{4}$ & $5\times10^{4}$ & $10^{5}$
\\
\hline
CO      & 1       &          $2.11\times10^{ 6}$ &          $1.22\times10^{ 8}$ &          $3.04\times10^{10}$ & $\mathbf{1.20\times10^{11}}$ & $\mathbf{2.31\times10^{11}}$
\\
CO      & 5       &          $2.03\times10^{ 7}$ &          $7.89\times10^{10}$ & $\mathbf{1.42\times10^{11}}$ & $\mathbf{3.35\times10^{11}}$ & $\mathbf{1.26\times10^{12}}$
\\
CO      & 10      &          $4.72\times10^{ 7}$ &          $8.52\times10^{10}$ & $\mathbf{1.57\times10^{11}}$ & $\mathbf{6.86\times10^{11}}$ & $\mathbf{2.56\times10^{12}}$
\\
CO      & 50      &          $4.32\times10^{ 8}$ &          $9.93\times10^{10}$ & $\mathbf{2.87\times10^{11}}$ & $\mathbf{2.87\times10^{12}}$ & $\mathbf{1.15\times10^{13}}$
\\
CO      & 100     &          $9.54\times10^{ 8}$ &          $3.31\times10^{10}$ & $\mathbf{2.62\times10^{11}}$ & $\mathbf{5.00\times10^{12}}$ & $\mathbf{2.04\times10^{13}}$
\\
\hline
CH      & 1       &          $7.51\times10^{ 8}$ &          $1.35\times10^{10}$ & $\mathbf{3.74\times10^{11}}$ & $\mathbf{6.49\times10^{11}}$ & $\mathbf{1.14\times10^{12}}$
\\
CH      & 5       &          $1.84\times10^{ 9}$ & $\mathbf{2.28\times10^{11}}$ & $\mathbf{4.46\times10^{11}}$ & $\mathbf{4.79\times10^{11}}$ & $\mathbf{1.13\times10^{12}}$
\\
CH      & 10      &          $2.12\times10^{ 9}$ & $\mathbf{1.37\times10^{11}}$ & $\mathbf{3.06\times10^{11}}$ & $\mathbf{4.52\times10^{11}}$ & $\mathbf{1.02\times10^{12}}$
\\
CH      & 50      &          $2.84\times10^{ 9}$ &          $2.87\times10^{10}$ & $\mathbf{1.18\times10^{11}}$ & $\mathbf{3.24\times10^{11}}$ & $\mathbf{7.35\times10^{11}}$
\\
CH      & 100     &          $2.82\times10^{ 9}$ &          $1.51\times10^{10}$ &          $5.81\times10^{10}$ & $\mathbf{2.77\times10^{11}}$ & $\mathbf{6.04\times10^{11}}$
\\
\hline
C$_2$   & 1       &          $2.20\times10^{ 5}$ &          $3.05\times10^{ 7}$ &          $1.71\times10^{ 9}$ &          $1.86\times10^{10}$ &          $7.97\times10^{10}$
\\
C$_2$   & 5       &          $4.32\times10^{ 6}$ &          $4.89\times10^{ 9}$ &          $1.11\times10^{10}$ &          $8.77\times10^{10}$ & $\mathbf{4.71\times10^{11}}$
\\
C$_2$   & 10      &          $1.07\times10^{ 7}$ &          $5.76\times10^{ 9}$ &          $1.49\times10^{10}$ & $\mathbf{1.79\times10^{11}}$ & $\mathbf{8.78\times10^{11}}$
\\
C$_2$   & 50      &          $8.52\times10^{ 7}$ &          $7.36\times10^{ 9}$ &          $3.64\times10^{10}$ & $\mathbf{5.50\times10^{11}}$ & $\mathbf{1.77\times10^{12}}$
\\
C$_2$   & 100     &          $1.66\times10^{ 8}$ &          $5.07\times10^{ 9}$ &          $4.01\times10^{10}$ & $\mathbf{6.09\times10^{11}}$ & $\mathbf{1.47\times10^{12}}$
\\
\hline
OH      & 1       &          $3.23\times10^{ 7}$ &          $4.92\times10^{ 8}$ & $\mathbf{9.81\times10^{11}}$ & $\mathbf{6.58\times10^{11}}$ & $\mathbf{2.85\times10^{11}}$
\\
OH      & 5       &          $1.56\times10^{ 8}$ & $\mathbf{8.13\times10^{11}}$ & $\mathbf{1.05\times10^{12}}$ & $\mathbf{2.96\times10^{11}}$ & $\mathbf{4.10\times10^{11}}$
\\
OH      & 10      &          $3.01\times10^{ 8}$ & $\mathbf{4.97\times10^{11}}$ & $\mathbf{6.58\times10^{11}}$ & $\mathbf{2.95\times10^{11}}$ & $\mathbf{3.28\times10^{11}}$
\\
OH      & 50      &          $1.43\times10^{ 9}$ & $\mathbf{1.25\times10^{11}}$ & $\mathbf{2.66\times10^{11}}$ & $\mathbf{1.72\times10^{11}}$ & $\mathbf{2.42\times10^{11}}$
\\
OH      & 100     &          $1.98\times10^{ 9}$ &          $1.51\times10^{10}$ &          $9.12\times10^{10}$ & $\mathbf{1.44\times10^{11}}$ & $\mathbf{2.42\times10^{11}}$
\\
\hline
\end{tabular}

 \end{center}
\end{table*}

\begin{table*}
 \caption{Column densities (cm$^{-2}$) for neutral and singly ionised
 calcium along the line of sight within a microstructure of transverse
 depth 10 AU and elongation factor 10 along the line of
 sight. The region is subject to standard interstellar radiation and
 particle fields.}
 \label{coldens2}
 \begin{center}
  \begin{tabular}{ccccccc}
\hline
Species & Time      & \multicolumn{5}{c}{Density (cm$^{-3}$)}
\\
        & (years)   & $10^{3}$ & $5\times10^{3}$ & $10^{4}$ & $5\times10^{4}$ & $10^{5}$
\\
\hline
Ca\textsc{I}  & 1   & $5.62\times10^{ 4}$ & $1.90\times10^{ 6}$ & $7.49\times10^{ 6}$ & $1.80\times10^{ 8}$ & $6.41\times10^{ 8}$
\\
Ca\textsc{I}  & 5   & $2.75\times10^{ 5}$ & $9.28\times10^{ 6}$ & $3.63\times10^{ 7}$ & $8.20\times10^{ 8}$ & $2.73\times10^{ 9}$
\\
Ca\textsc{I}  & 10  & $5.39\times10^{ 5}$ & $1.81\times10^{ 7}$ & $6.98\times10^{ 7}$ & $1.46\times10^{ 9}$ & $4.43\times10^{ 9}$
\\
Ca\textsc{I}  & 50  & $2.36\times10^{ 6}$ & $7.38\times10^{ 7}$ & $2.64\times10^{ 8}$ & $3.19\times10^{ 9}$ & $6.26\times10^{ 9}$
\\
Ca\textsc{I}  & 100 & $4.00\times10^{ 6}$ & $1.17\times10^{ 8}$ & $3.82\times10^{ 8}$ & $3.12\times10^{ 9}$ & $5.13\times10^{ 9}$
\\
\hline
Ca\textsc{II} & 1   & $1.10\times10^{ 9}$ & $4.30\times10^{ 9}$ & $1.02\times10^{10}$ & $5.50\times10^{10}$ & $1.13\times10^{11}$
\\
Ca\textsc{II} & 5   & $1.10\times10^{ 9}$ & $4.30\times10^{ 9}$ & $1.02\times10^{10}$ & $5.44\times10^{10}$ & $1.11\times10^{11}$
\\
Ca\textsc{II} & 10  & $1.17\times10^{ 9}$ & $4.82\times10^{ 9}$ & $9.39\times10^{ 9}$ & $5.38\times10^{10}$ & $1.01\times10^{11}$
\\
Ca\textsc{II} & 50  & $1.16\times10^{ 9}$ & $4.76\times10^{ 9}$ & $9.20\times10^{ 9}$ & $5.19\times10^{10}$ & $9.95\times10^{10}$
\\
Ca\textsc{II} & 100 & $1.16\times10^{ 9}$ & $4.72\times10^{ 9}$ & $9.08\times10^{ 9}$ & $5.17\times10^{10}$ & $1.01\times10^{11}$
\\
\hline
\end{tabular}

 \end{center}
\end{table*}

\begin{figure}
 \caption{Contours of log CO column density as a function of density
 and time for a microstructure of transverse depth 10 AU,
 elongated by a factor of 10 along the line of sight, and a cosmic ray
 ionisation rate of $\zeta=1.3$$\times$10$^{-17}$ s$^{-1}$. From top
 to bottom, the three figures are for incident radiation field
 strengths of $1/3$, 1 and 3 Habing, respectively.}
 \label{coplot}
 \begin{center}
  \includegraphics[width=0.45\textwidth]{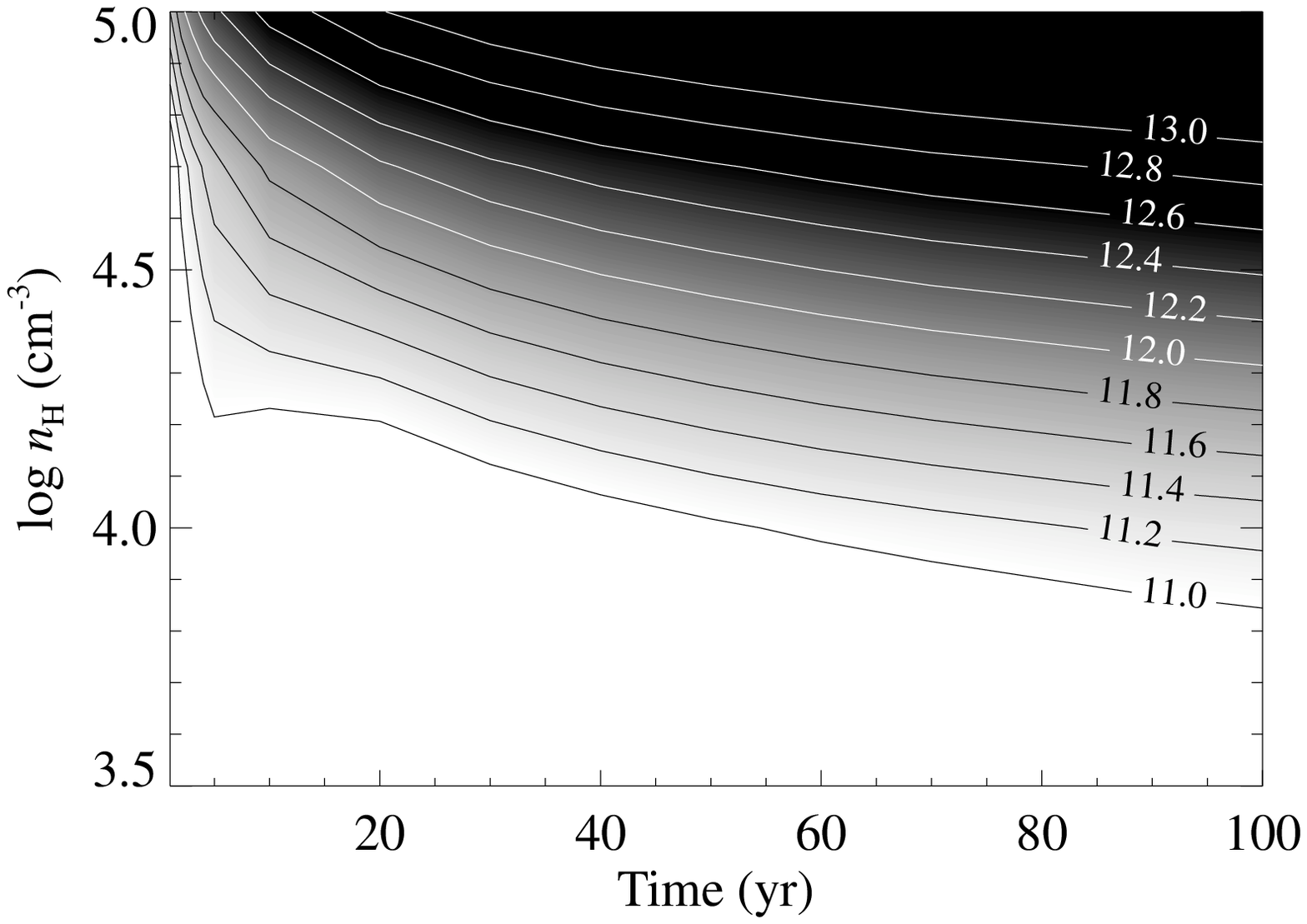}
  \includegraphics[width=0.45\textwidth]{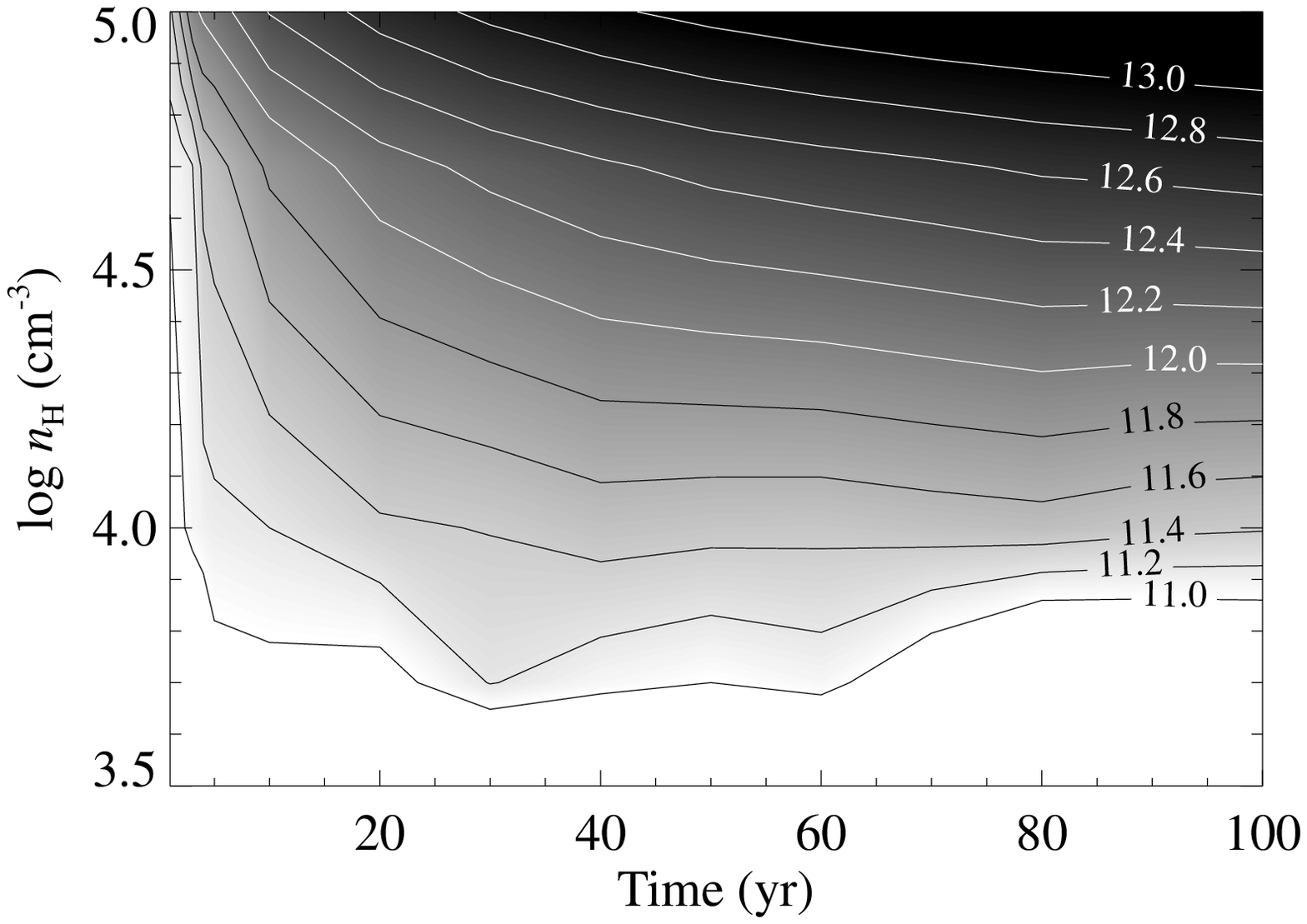}
  \includegraphics[width=0.45\textwidth]{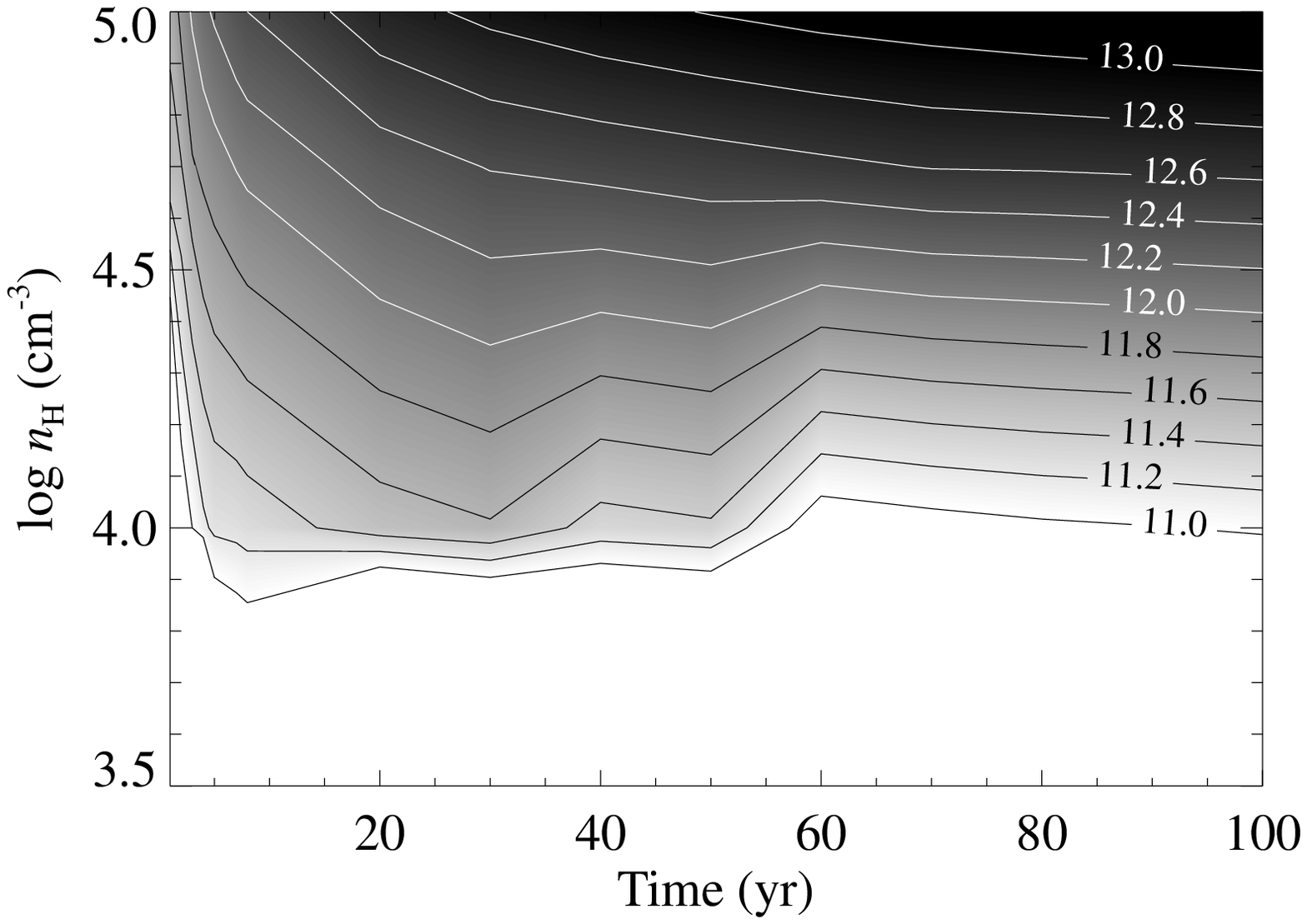}
 \end{center}
\end{figure}

\begin{figure}
 \caption{Contours of log CH column density as a function of density
 and time for a microstructure of transverse depth 10 AU,
 elongated by a factor of 10 along the line of sight, and a cosmic ray
 ionisation rate of $\zeta=1.3$$\times$10$^{-17}$ s$^{-1}$. From top
 to bottom, the three figures are for incident radiation field
 strengths of $1/3$, 1 and 3 Habing, respectively.}
 \label{chplot}
 \begin{center}
  \includegraphics[width=0.45\textwidth]{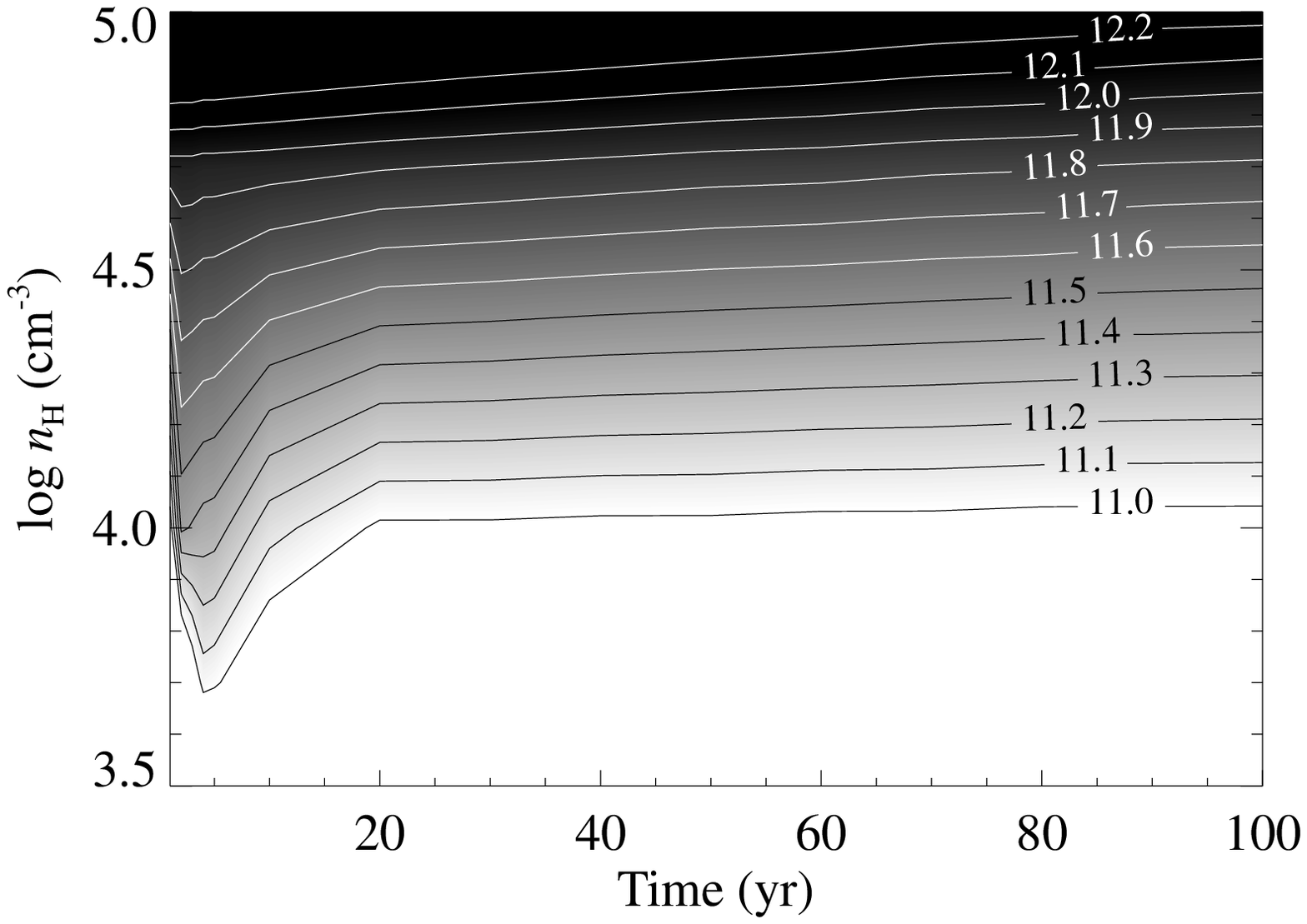}
  \includegraphics[width=0.45\textwidth]{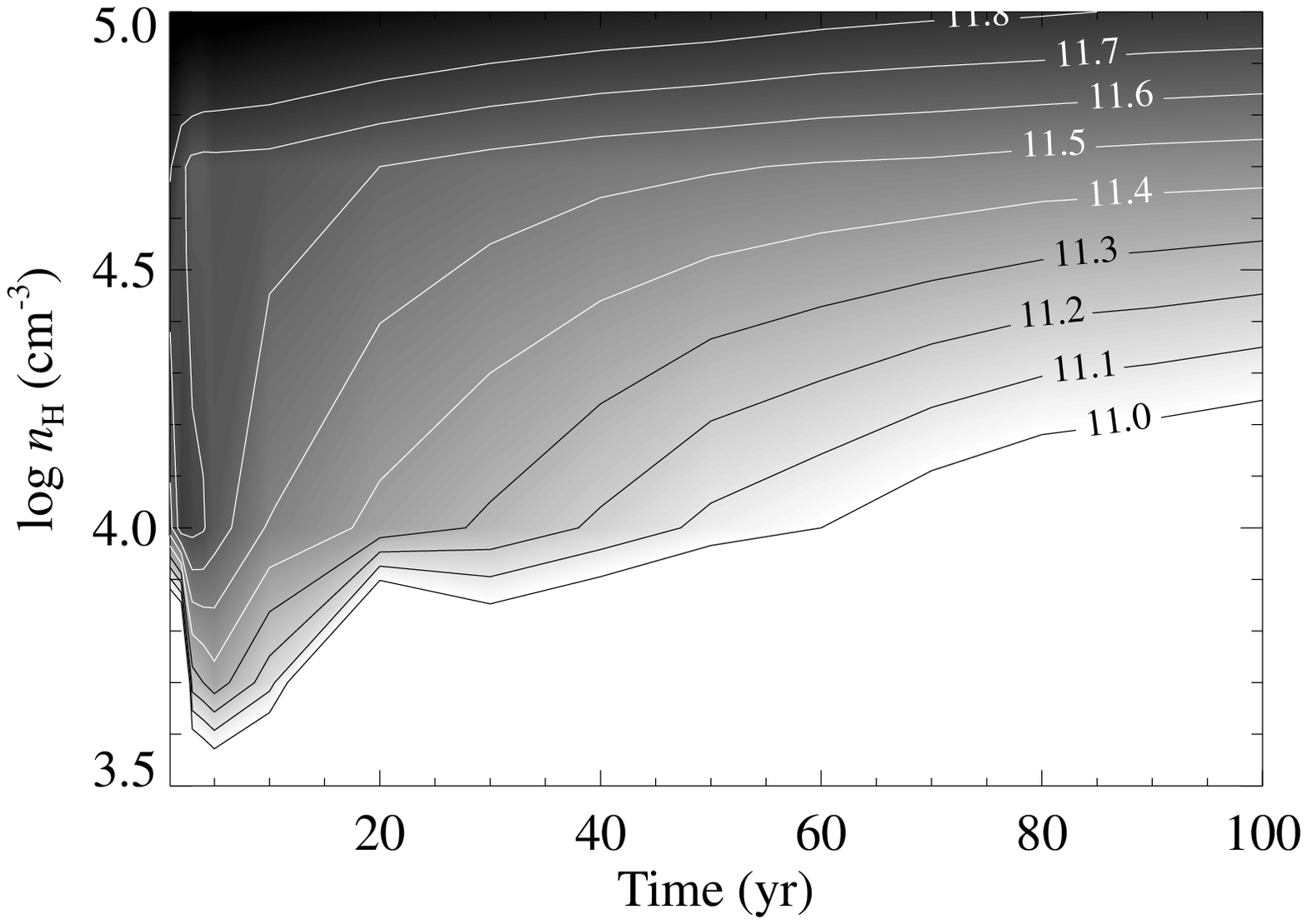}
  \includegraphics[width=0.45\textwidth]{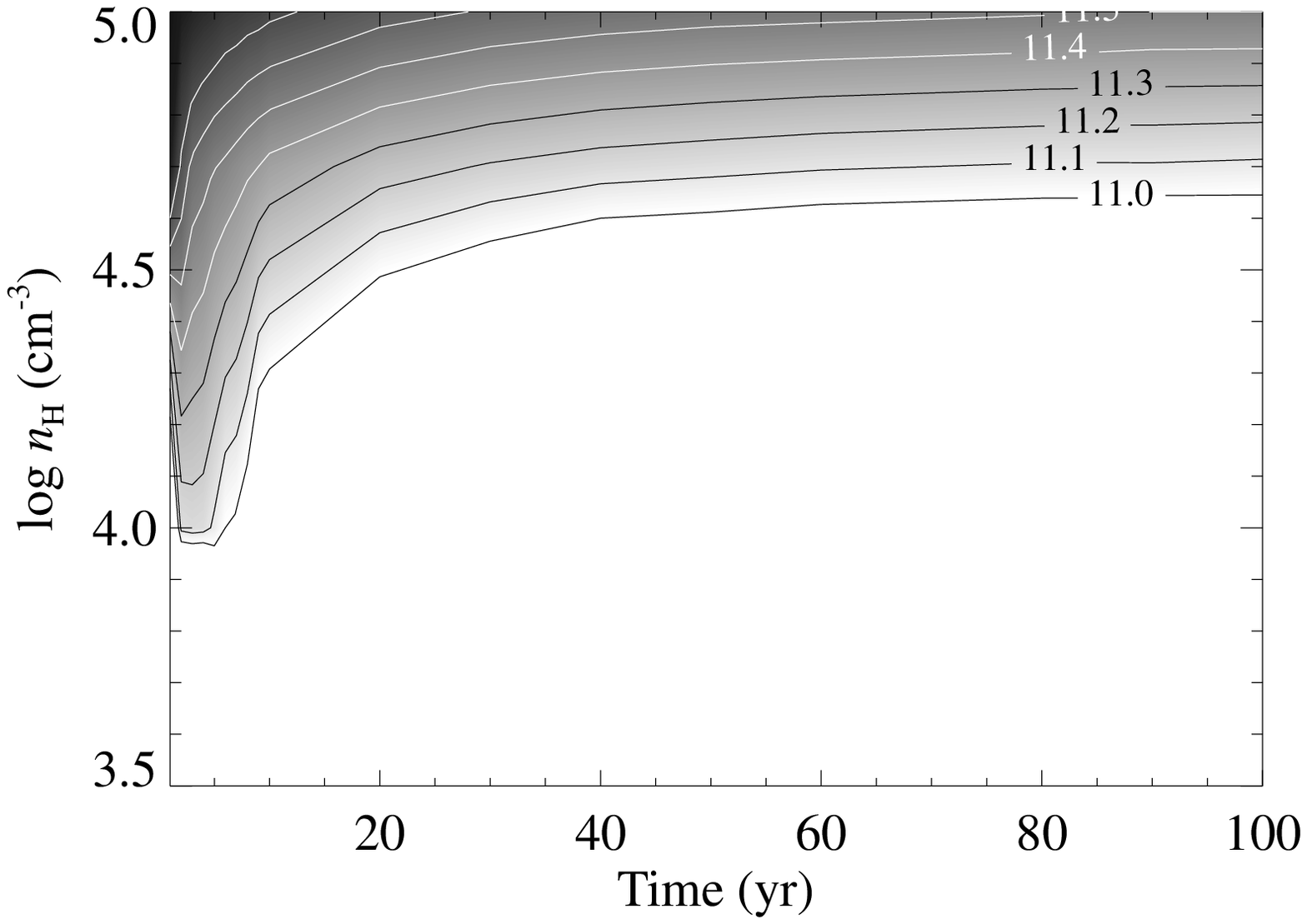}
 \end{center}
\end{figure}

\begin{figure}
 \caption{Fractional abundances of CO (solid line), CH (dotted), C$_2$
 (dashed) and OH (dash-dot) at a transverse visual extinction
 of $A_{V,trans}=10^{-4}$ mags into the microstructure, subject to an
 incident radiation field strength of 1 Habing and a cosmic ray
 ionisation rate of $\zeta=1.3$$\times$10$^{-17}$ s$^{-1}$. From top
 to bottom, the three figures are for microstructure densities of
 10$^3$, 10$^4$ and 10$^5$ cm$^{-3}$, respectively.}
 \label{abundances}
 \begin{center}
  \includegraphics[width=0.45\textwidth]{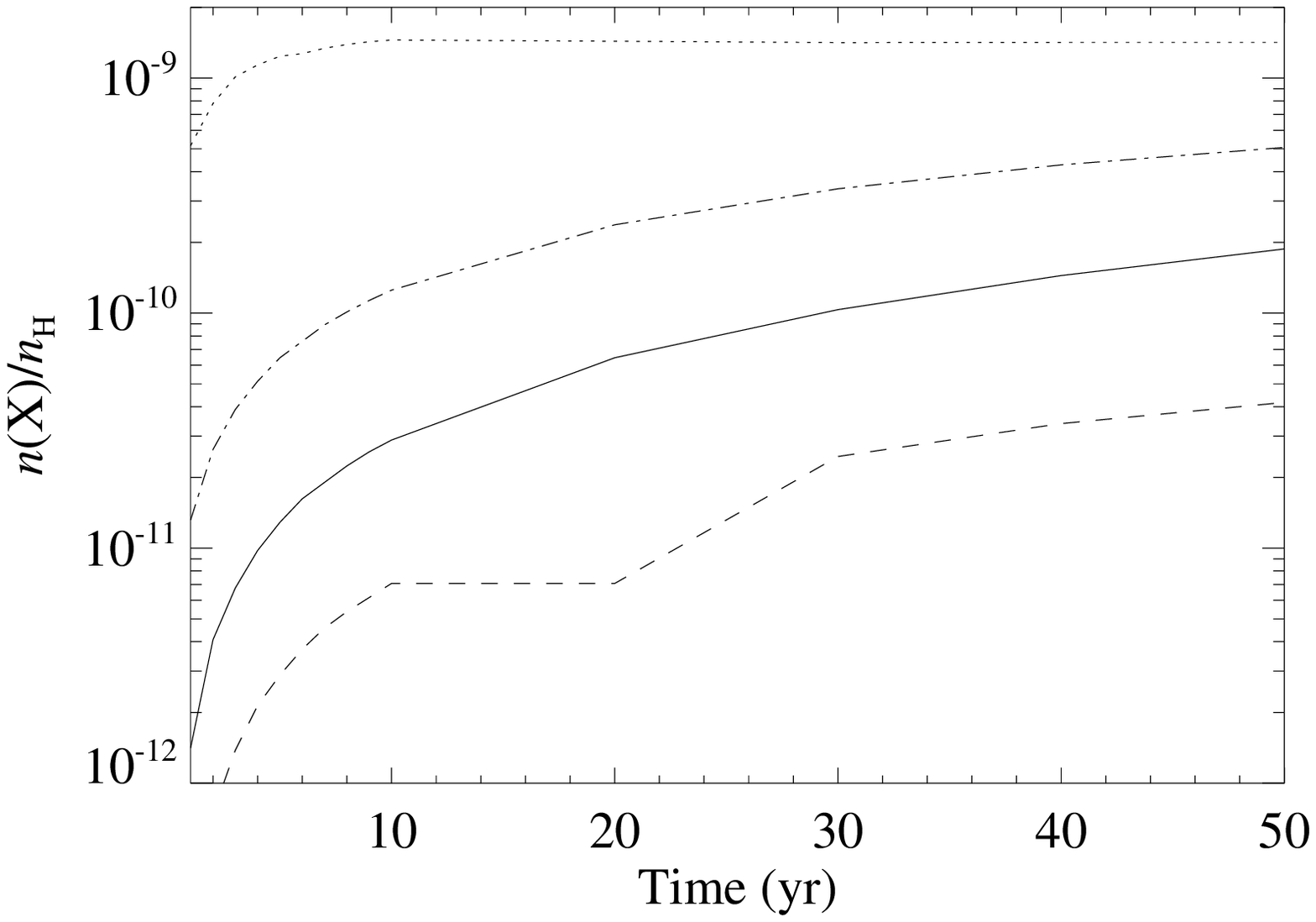}
  \includegraphics[width=0.45\textwidth]{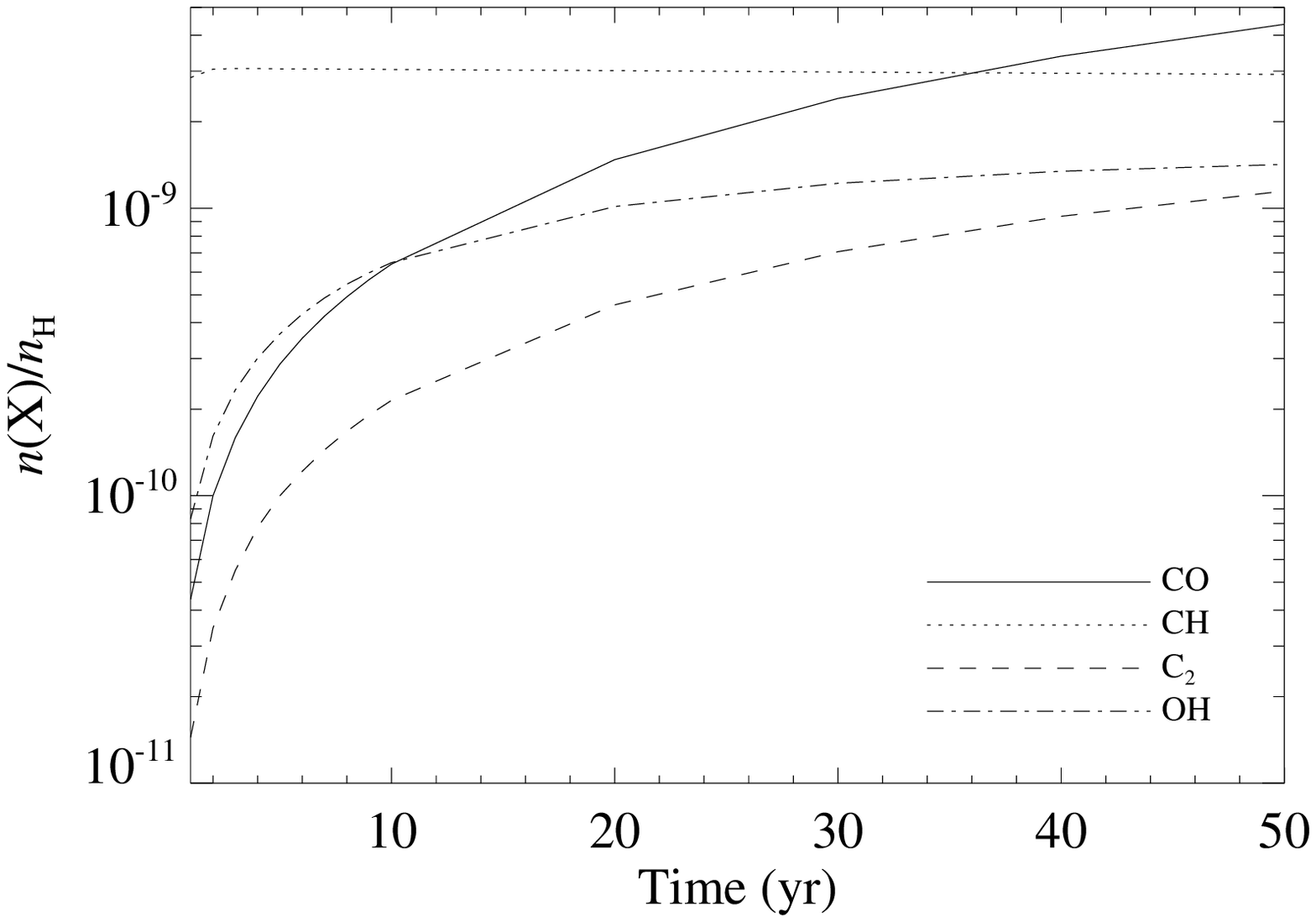}
  \includegraphics[width=0.45\textwidth]{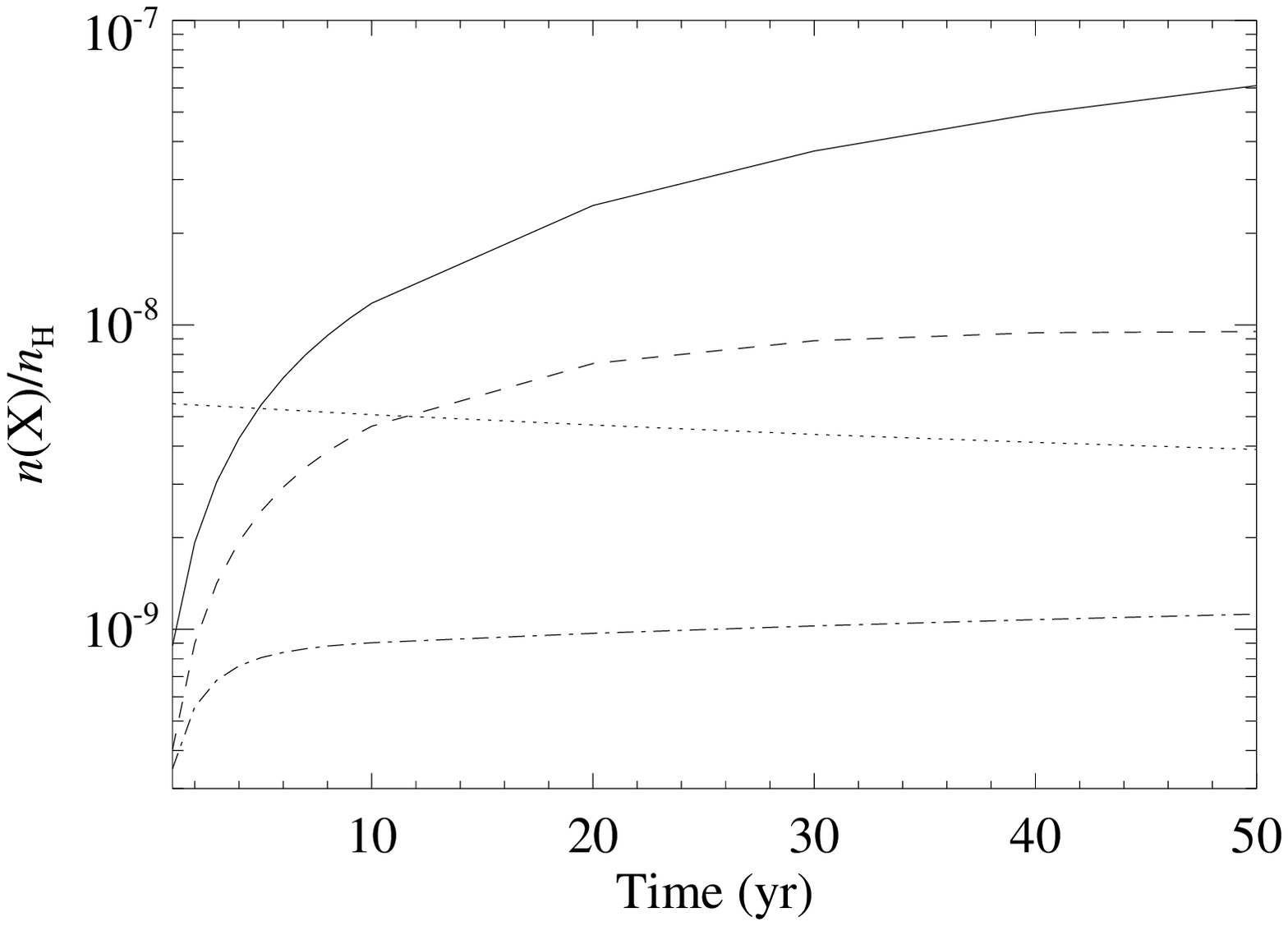}
 \end{center}
\end{figure}

The purpose of our study is to investigate whether there exists a
parameter space of densities, radiation fields, cosmic ray ionisation
rates and dimensions where molecules can be formed in the variable
components, and also to determine the time-dependent nature of such
structures. In the next two subsections we make predictions, based on
our models, of the physical conditions and time-dependence of the
microstructure, respectively.

In all the results reported here, we have assumed an elongation factor
of 10; i.e. the path-length along the line of sight through the
microstructure is 10$\times$ the transverse depth, 10 AU.

\subsection{Physical Characteristics of the Microstructure}
The results of our chemical model suggest that only a few
molecular species are formed in large enough quantities in
the proposed scenario to be potentially detectable. They include CH,
CH$_2$, CO, C$_2$ and OH. In our model, these species are only
produced in great enough quantities to be detectable under certain
physical conditions (see Tables~\ref{coldens1} and
\ref{coldens2}, and Figures~\ref{coplot} and \ref{chplot}). In
Table~\ref{coldens1} we list the column densities of CO, CH, C$_2$ and
OH as a function of density and time at a transverse depth of
10 AU into the microstructure slab, corresponding to a
transverse visual extinction of approximately 10$^{-4}$ to
10$^{-2}$ mags for number densities in the range 10$^3$ to 10$^5$
cm$^{-3}$, making the assumption that the elongation factor
along the line of sight (described above) is 10. Column
densities larger than 10$^{11}$ cm$^{-2}$ are shown in bold in
Table~\ref{coldens1}, indicating a reasonable lower limit for
molecular species that are potentially detectable. Assuming that the
microstructure has a transverse spatial dimension of 10 AU, a
minimum density of 10$^4$ cm$^{-3}$ is necessary for most species to
be detectable. Table~\ref{coldens2} lists the column densities
for Ca\textsc{I} and Ca\textsc{II}. Since atoms have stronger
oscillator strengths than molecules, these species are detectable at
lower column densities than the limit of 10$^{11}$ cm$^{-2}$ imposed
in Table~\ref{coldens1}. Figures~\ref{coplot} and \ref{chplot} show
how the column densities of CO and CH behave as a function of density,
time and radiation field. We find that CO is particularly sensitive to
variations in the number density, and its column density can vary by
over 2 orders of magnitude in the number density range
considered here. Its sensitivity to the radiation field is more
subtle: at early times ($<20$ yr) and low densities ($<10^4$
cm$^{-3}$), a weak radiation field ($<1$ Habing) yields a longer
timescale ($\sim100$ yr) for the production of detectable CO. CH, on
the other hand, shows a strong sensitivity to the radiation field
employed. In fact, if the field is stronger than 1 Habing, CH can only
remain detectable at high densities ($>5$$\times$10$^4$ cm$^{-3}$). In
general, CH is formed and then destroyed very quickly. These trends
restrict the parameter space for which both CO and CH can be present
in the microstructure \textit{at the same time}. None of the
species seem to have a strong dependence on the rate of cosmic ray
ionisation employed.

The most remarkable result is that, even at the earliest times shown,
if the number density is sufficiently high then potentially detectable
column densities of the species listed above are attained. Thus, the
high density drives a chemistry fast enough to overcome the losses due
to the strong radiation field. Since the transverse visual
extinctions considered are close to zero, the region is essentially
unshielded. Note, e.g., that the total extinction along the line of
sight toward $\kappa$ Vel observed by \citet{b5} is small
($E(B-V)=0.11$) and that the transverse extinction at the
variable component must be significantly less than this. By assuming
that the transverse extinction is close to zero, we are
therefore estimating the lower limit of the additional contribution to
column densities from species generated in the microstructure.

The adopted column density of H$_2$ is, however, sufficiently large to
ensure that the H$_2$ photodissociation rate is too slow to cause
significant loss of H$_2$ on the timescales considered here.

The fact that high densities are necessary in order to produce
sufficiently large molecular abundances in the microstructure
is consistent with the conclusions of \citet{b31} who find that, if
the molecular variability is indeed due to density structure, the
number densities need to be at least 10$^6$ cm$^{-3}$. In fact,
Marscher et al.~(1993) and \citet{b31} detect both H$_2$CO and OH in
their time variability study. In our study, OH is present as long as
the density is $>5$$\times$10$^3$ cm$^{-3}$ and does indeed show time
variability (see Table~\ref{coldens1}). H$_2$CO, however, never
reaches a high enough abundance to be detectable for the number
densities considered here. In order to see whether densities as high
as the ones considered by \citet{b31} could produce significant
amounts of H$_2$CO, we ran two more models at 10$^6$ and 10$^7$
cm$^{-3}$ under standard interstellar conditions. We find that at
10$^6$ cm$^{-3}$ H$_2$CO is only just barely close to the lower limit
of detectability for times $\ge500$ yr, while at 10$^7$ cm$^{-3}$,
H$_2$CO becomes detectable after only 50 years. These models further
support the \citet{b31} findings.

\subsection{Time-Dependence of the Chemistry}
Tables~\ref{coldens1} and \ref{coldens2}, and
Figure~\ref{abundances} show how the different species behave as a
function of time. We now briefly discuss the time variability of each
species for number densities $\ge10^4$ cm$^{-3}$. CO takes $\sim$ 5 --
10 years to form in significant amounts, and its abundance increases
steadily with time. For CO to be abundant, therefore, the
microstructure would need to exist for at least 10 years. CH, on the
other hand, forms in less than 1 year, but is subsequently destroyed
(in part to form CO) within the first 100 years. C$_2$ takes over 10
years to form in detectable quantities, then survives in the gas for
at least 300 years, although its abundance starts to decline. OH
behaves in a similar manner to CH in that it forms within 1 year but
starts being converted into other species almost immediately and by
500 yr its abundance is too low to be detectable.
Ca\textsc{II} is present, confirming the detection of
\citet{b5} at the velocity of the microstructure. However, it cannot
be used as a diagnostic test of the microstructure age because its
abundance is invariant. Ca\textsc{I}, on the other hand, shows
considerable time variability, though whether the variations are large
enough to be detectable is less certain. Of course, these variations
are not reflected in the column densities of Ca\textsc{II}, since its
abundance is several orders of magnitude higher. The CO column
density under appropriate parameter choices generally exceeds those of
other molecular species. Thus, for a number density of 10$^4$
cm$^{-3}$, by 100 yr the CO column density has built up to
2.6$\times$10$^{11}$ cm$^{-2}$, while those for CH and C$_2$ are
5.8$\times$10$^{10}$ and 4.0$\times$10$^{10}$ cm$^{-2}$
respectively. Although the amounts by which CH, OH, C$_2$,
etc. decline with time are very small and may therefore seem
insignificant, they are indeed real and consistent with the gas phase
chemistry that is taking place. The behaviour described above is
expected for the usual gas phase chemical networks: CO and C$_2$
column densities increase monotonically over the time span considered
here. A feed molecule for the networks producing CO and C$_2$ is CH,
and its column density declines slightly over this period, showing
that CH is being converted into other species; no significant loss of
CH by photodissociation occurs in the time interval explored here.

In summary, for CO and C$_2$ to form, the microstructure needs to
survive at least 10 years. On the other hand, it cannot be much older
than $\sim50$ yr if CH is indeed observed.

Unfortunately, detection of CO by absorption in its resonance line
requires a UV spectrometer in an orbital observatory; none is
currently available. However, CH and C$_2$ are detectable from the
ground. It appears from this work that these molecules should have
comparable abundances in transient microstructure of number density
$\sim10^4$ cm$^{-3}$.

\section{Comparisons with Observations and Conclusions}
This study was prompted in part by the molecular detection obtained by
\citet{b5} at the velocity (different to that of the
surrounding medium) of previously reported microstructure
in the line of sight toward $\kappa$ Vel. Hence we now briefly
compare our models with such observations. In Table~2 of \citet{b5},
the observed column densities of Ca\textsc{I}, Ca\textsc{II} (two
velocity components) and CH are listed as, respectively,
3.2$\times$10$^9$, 2.0$\times$10$^{11}$ and 6.2$\times$10$^{10}$,
2.7$\times$10$^{11}$ cm$^{-2}$. If we assume an elongation factor of
10 along the line of sight and a gas density of the order of
$\sim10^5$ cm$^{-3}$, the observed CH column density can be easily
reached at very early times ($\sim10$ yr) and maintained. If instead
the density of the microstructure is closer to 10$^4$ cm$^{-3}$ then
we infer a maximum age for the microstructure of about 50 yr, after
which most CH would be destroyed. Hence observations of CH indicate
that such filamentary or sheet-like structures are likely to
be younger than 50 yr. The observed Ca\textsc{I} and
Ca\textsc{II} column densities can be produced in the model if the
microstructure has a number density of $\sim5$$\times$10$^4$ cm$^{-3}$
and is older than 10 yr (see Table~\ref{coldens2}).

In order to further test the predicted molecular abundances, we
conducted a high-resolution search for interstellar C$_2$ at the
velocity of the variable component toward $\kappa$ Vel using the
Ultra-High-Resolution Facility (UHRF; Diego et al. 1995) at the AAT in
February 2003. Sixteen individual exposures were made of the region
containing the Q(4) line at 8763.751~\AA, with a total integration
time of 4.9 hours. Data reduction and calibration was as described by
\citet{b4,b5}. The final continuum signal-to-noise ratio was 480, but
the Q(4) line was not detected. The resulting 3$\sigma$ upper limit to
the equivalent width of the Q(4) line was 0.09~m\AA, yielding an upper
limit to the $J=4$ column density of $N(J=4)\leq1.95$$\times$10$^{11}$
cm$^{-2}$ (using the oscillator strength adopted by Crawford 1997). At
the high densities ($n_\mathrm{H}=10^3$ to 10$^5$ cm$^{-3}$) expected
for these variable components, the rotational populations will be
approximately thermal, and for $T\sim100$~K we expect the total C$_2$
column density to be $N(\mathrm{C_2})\sim4\times N(J=4)$ (e.g. van
Dishoeck 1984). Thus, our observations imply $N(\mathrm{C_2})\lesssim
10^{12}$ cm$^{-2}$ in the variable interstellar component toward
$\kappa$ Vel. This is consistent with the model results given in
Table~\ref{coldens1} for $N_{\mathrm{H}}\lesssim10^5$ cm$^{-3}$. More
sensitive searches for C$_2$ would appear to be worthwhile, as these
would be able to constrain the model results at lower densities.

The best matching models for observations at the velocity of
the variable component in the $\kappa$ Vel line of sight seem
therefore to be those for a young ($\sim50$ yr), transient ($<100$
yr), high density ($>10^4$ cm$^{-3}$), small ($\sim10$ AU), slightly
elongated (by about a factor of 10 along the line of sight) structure,
immersed in an ambient radiation field.

The main result of this paper is that essentially unshielded interstellar
gas can develop detectable column densities of molecules on very short
time-scales ($<100$ yr) for path-lengths on the order of 100 AU
along the line of sight, if the density is $\sim10^4$
cm$^{-3}$ or higher. This time and density domain is a region of
parameter space not previously explored. The calculations presented
here are pseudo time-dependent, in that the slab structure is fixed. 

If the cause of such microstructure is indeed magnetohydrodynamic,
then the filamentary objects likely to be responsible for the effects
observed toward $\kappa$ Vel and elsewhere are probably waves moving
at magnetosonic sound speed. More complex and realistic models are now
being explored \citep{b14}.

\section*{Acknowledgments}
TAB is supported by a PPARC studentship. SV acknowledges individual
financial support from a PPARC Advanced Fellowship. DAW thanks the
Leverhulme Fund for the Leverhulme Emeritus Award. The authors are
very grateful to Dr T.~Nyugen for initial studies of these
problems. We thank the referee for constructive comments which helped
to improve an earlier draft of this paper.

\bsp

\label{lastpage}

\end{document}